\documentclass[aps,prb,twocolumn]{revtex4-1}

\usepackage{graphicx}
\usepackage{dcolumn}
\usepackage{bm}
\usepackage{amsmath}
\usepackage{textcomp}
\usepackage{subcaption}
\usepackage{ulem}
\usepackage{epsfig}
\usepackage{epstopdf}
\usepackage{float}
\usepackage{color}

\begin{document}

\title{Monte Carlo simulations of the kagome lattice with magnetic dipolar interactions.}


\author{M. S. Holden}
\affiliation{Department of Physics and Physical Oceanography, Memorial University of Newfoundland,
St. John's, Newfoundland, A1B 3X7,  Canada}
\author{M. L. Plumer}
\affiliation{Department of Physics and Physical Oceanography, Memorial University of Newfoundland,
St. John's, Newfoundland, A1B 3X7,  Canada}
\author{I. Saika-Voivod}
\affiliation{Department of Physics and Physical Oceanography, Memorial University of Newfoundland,
St. John's, Newfoundland, A1B 3X7,  Canada}
\author{B. W. Southern}
\affiliation{Department of Physics and Astronomy, University of Manitoba, Winnipeg, MB, R3T 2N2,
Canada}

\date{\today}

\begin{abstract}
The results of extensive Monte Carlo simulations of classical spins on the two-dimensional kagome lattice with only dipolar interactions are presented.   In addition to revealing the six-fold degenerate ground state, the nature of the finite-temperature phase transition to long-range magnetic order is discussed.  Low temperature states consisting of mixtures of degenerate ground state configurations separated by domain walls can be explained as a result of competing exchange-like and shape anisotropy-like terms in the dipolar coupling. Fluctuations between pairs of degenerate spin configurations are found to persist well into the ordered state as the temperature is lowered until locking in to a low-energy state.  
\end{abstract}


\maketitle
\section{INTRODUCTION}
The study of spin systems with short-range antiferromagnetic exchange interactions on geometrically frustrated lattices has 
revealed a remarkably wide variety of magnetic structures and phase transitions.\cite{frustration}
Exotic spin states resulting from frustration through the lattice geometry or magnetic interactions couple
to other degrees of freedom and are associated with a range of disparate phenomena including 
anomalous hall effect,\cite{chen14} the magnetoelectric effect\cite{magneto} and exchange-biasing in spin valves.\cite{tsunoda10,aley11}
The study of magnetic dipole-dipole interactions on lattice types which are known to induce frustration with only near-neighbor 
exchange coupling have recieved much less attention due to the relative weakness of magnetostatic effects. An exception
is the pyrochlore latice and its association with spin ice phenomena.\cite{bramwell} Recent attention has also been devoted to artifically templated magnetic islands on frustrated and unfrustrated lattice types which are well separated and interact only through magnetostatic coupling.\cite{chioar14,xie14}

Although the dipole-dipole interaction is much smaller than exchange effects, its long-range nature can lead to significant
effects on magnetic structures in systems with ferromagnetic exchange.\cite{johnw} In cases where 
the fundamental magnetic interaction is antiferromagnetic, dipole effects tend to be much weaker 
due to smaller net magnetization.  However, for low-dimensional systems and at surfaces, there 
can be larger net magnetization in systems with antiferromagnetic exchange and dipole effects can be more improtant.  In 
spin-valve structures, the antiferromagnet provides exchange biasing to the adjacent thin 
ferromagnetic film, which exchange couples to it.\cite{ogrady}  The impact of dipole interactions 
within the surface layer of the antiferromagnet in this case may be an important effect for the 
explanation of the exchange-bias field but the phenomenon of exchange bias is not well understood. 
The most popular compound for use as the antiferromagnet in spin valves for magnetic recording is 
IrMn$_3$, which in its ordered crystalline fcc phase is composed of ABC stacked [111] planes of 
magnetic Mn ions on kagome sites.\cite{tsunoda10,aley11,kren66,tomeno,hemmati,leblanc1,leblanc2}

The large number of studies over the past few decades on the near-neighbor antiferromagnetic exchange Heisenberg kagome model have
shown that the large degeneracy associated with the 120$^\circ$ spin structures on corner-sharing 
triangles leads to
the stability of so-called $q=0$, $\sqrt{3} \times\sqrt{3}$ and other more exotic spin structures.\cite{kagome,harris,zhito,schnabel} 
This local 120$^\circ$ ordering also occurs in the edge-sharing triangular lattice with 
near-neighbor exchange.\cite{frustration} 
In the case of long-range dipole interactions, details of the lattice geometry can play a more important role.  For the
triangular lattice, the ground-state and low-temperature spin structures have been established to be ferromagentic whereas for the
kagome lattice, this appears to not be the case.\cite{mckeehan,politi,tomita} Spin-ice type order has been proposed for the kagome lattice using both Ising-like\cite{chern} as well as anisotropic Heisenberg\cite{wysin} spin models.
Of particular interest to the present work are the Monte Carlo (MC) simulation studies of Tomita\cite{tomita} on the kagome lattice using a purely Heisenberg spin model with dipole interactions.  These results suggest that alternating rows with six-fold ferromagnetic order, and disorder, are stabilized below a critical temperature of T$_N \simeq 0.43$ (in dimensionless units).  The ground state spin configuration was not studied.

In this work, extensive Metropolis MC simulations are perfomed on the two-dimensional kagome Heisenberg spin lattice with only dipole  interactions. We confirm a phase transition to long-range magnetic order but the spin configuration we find differs from that of Tomita.  
Significant fluctuations occur between coexisting six-fold non-ferromagnetic degenerate states over a range of temperature (T) before locking-in to a pair of states at lower T.  The nature of the ground state is examined using the effective field method and confirmed by low-T MC simulation results. The impact of the range of the interaction on the ground state is also examined.

The remainder of this paper is organized as follows. In Sec.II, the model is described and in Sec. 
III the effective field method (EFM) is used to reveal the six-fold degenerate spin order in the 
ground state.  MC simulations results for the magnetization, sub-lattice order parameters, specific 
heat and susceptibility are presented in Sec. IV.  A summary, conclusions and directions for future 
work are given in Sec. V.

\section{THE MODEL}
\label{sec-model}

The dimensionless dipole interaction Hamiltonian for Heisenberg model spins is given by:
\begin{equation} \label{hamilton}
 \mathcal{H}_{dip} = \sum_{i,j} \frac{\vec{S}_{i} \cdot \vec{S}_{j}}{r_{ij}^3} -
3\frac{(\vec{S}_{i} \cdot \vec{r}_{ij})(\vec{S}_{j} \cdot \vec{r}_{ij})}{r_{ij}^5}
\end{equation}
where $\vec{r}_{ij}$ is the (dimensionless) vector connecting spins on lattice sites $i$ and $j$. 
This Hamiltonian  contains two terms that offer some insight into
how the spins may tend to align at low temperatures. The second term in Eq.
\ref{hamilton} favours spins that align along the vectors $\vec{r}_{ij}$. Therefore, in the 
ground state it is reasonable to expect spins to exhibit some preferred orientation related to the 
lattice vectors. This phenomenon is referred to shape anisotropy and is dependent on the geometry
of the lattice. The first term is noticeably similar to the antiferromagnetic exchange interaction.

In view of what is known about the ground state of the exchange-only kagome
lattice, the ground state of the dipolar kagome system can be expected to be a three-sublattice 
system with fixed spin orientations for each sublattice. We
expect that in the ground state all spins on sites of the same sublattice will point in the same
direction, and so the ground state can be represented by three angles, one for each sublattice. From
the above arguments it can be seen that in the ground state, the spins will tend to lie in the plane
of the lattice due to the shape anisotropy induced by the dipole interaction.
In order to efficiently compute the long-range interaction, we use standard Ewald summation 
techniques.\cite{ewald} 

\section{THE GROUND STATE}
\label{sec:gs}

The EFM has been successful in the determination of complicated spin structures associated with spin glasses.\cite{walker}
In our implementation of this method, the starting point is an ordered or random initial spin 
configuration, the spins of which are iteratively aligned with the local field to minimize the 
energy.  This procedure is then repeated multiples times to ensure that the global minimum energy 
spin state is achieved. In addition, low temperature 
MC simulations 
presented in Section~\ref{sec:mc}
corroborate the ground state results.

\begin{figure}[H]
\centering
 \includegraphics[scale=0.55]{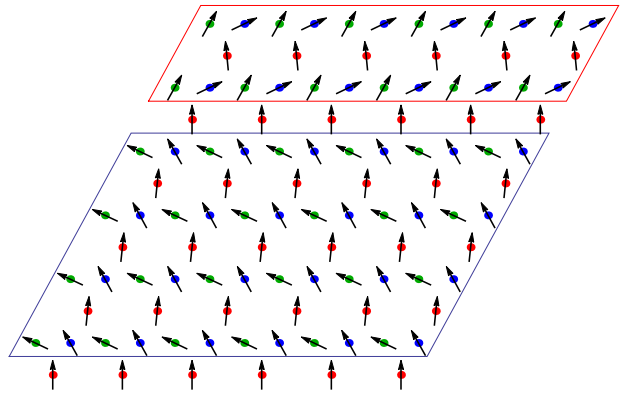}
 \caption[Low-temperature spin structure with distinct domains]{Example low-temperature spin structure
identified by MC and EFM simulations. The state contains states 1 \& 5 as defined in 
Table \ref{tabgs}. Domain walls separate the two ground states. Spins in a domain wall have an
orientation that is the average of the orientations of the spins belonging to the sublattice on
which the domain wall forms. The top two rows are state 1 while the bottom rows correspond to state 5.}
 \label{gssnap}
\end{figure}

\begin{figure}[H]
\centering
 \begin{subfigure}{0.23\textwidth}
  \includegraphics[width=\textwidth]{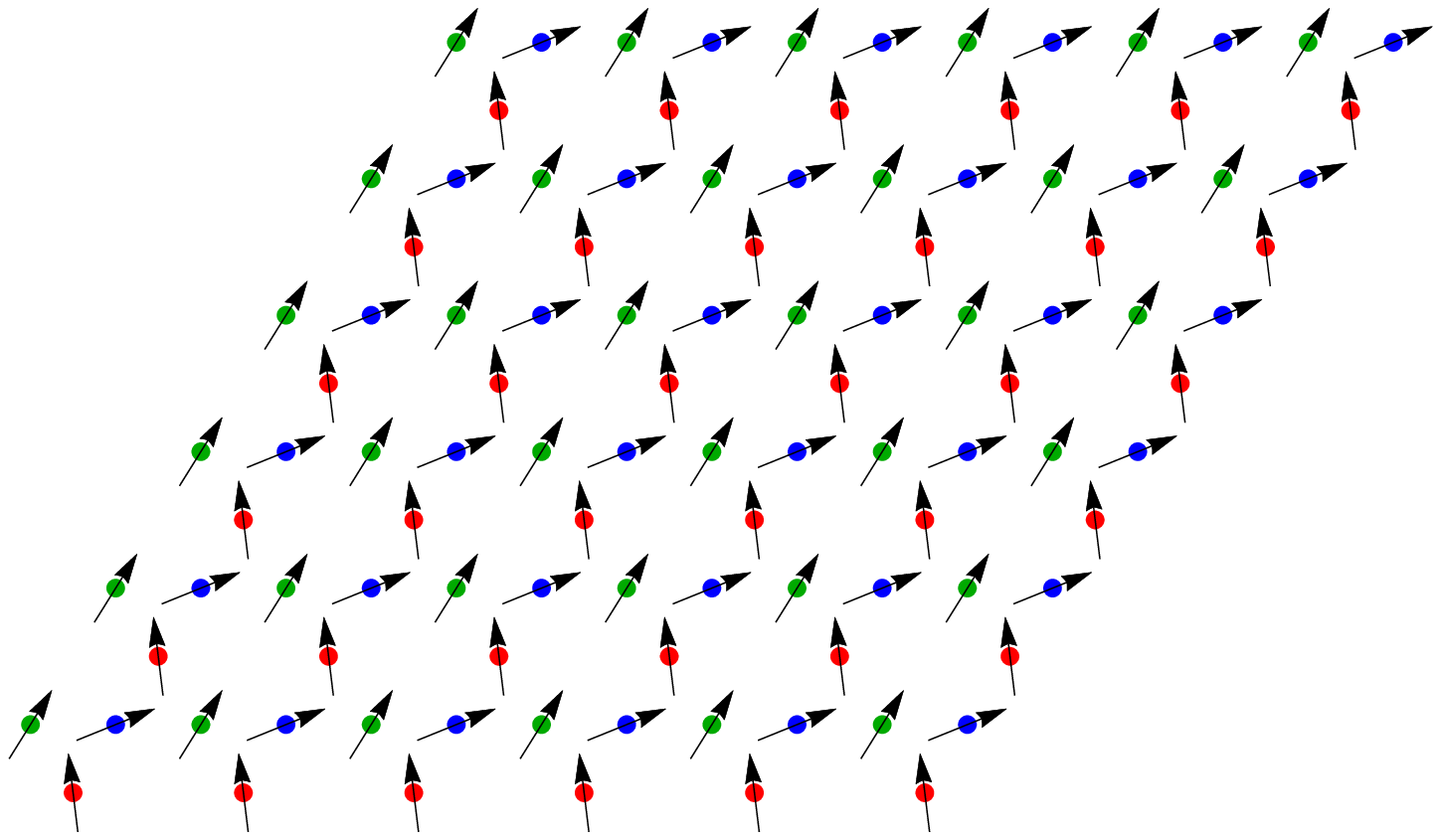}
  \caption{State 1}
 \end{subfigure}
 \begin{subfigure}{0.23\textwidth}
  \includegraphics[width=\textwidth]{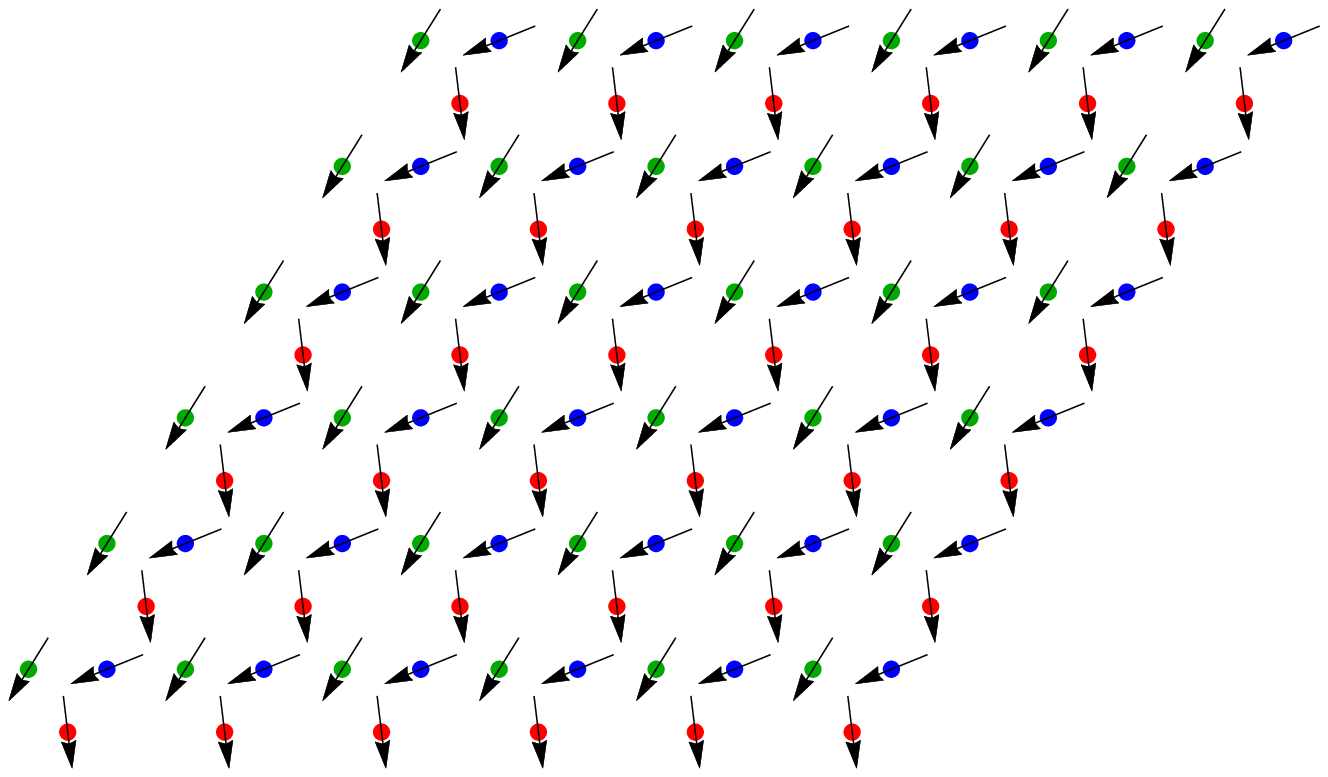}
  \caption{State 2}
 \end{subfigure}
 \begin{subfigure}{0.23\textwidth}
  \includegraphics[width=\textwidth]{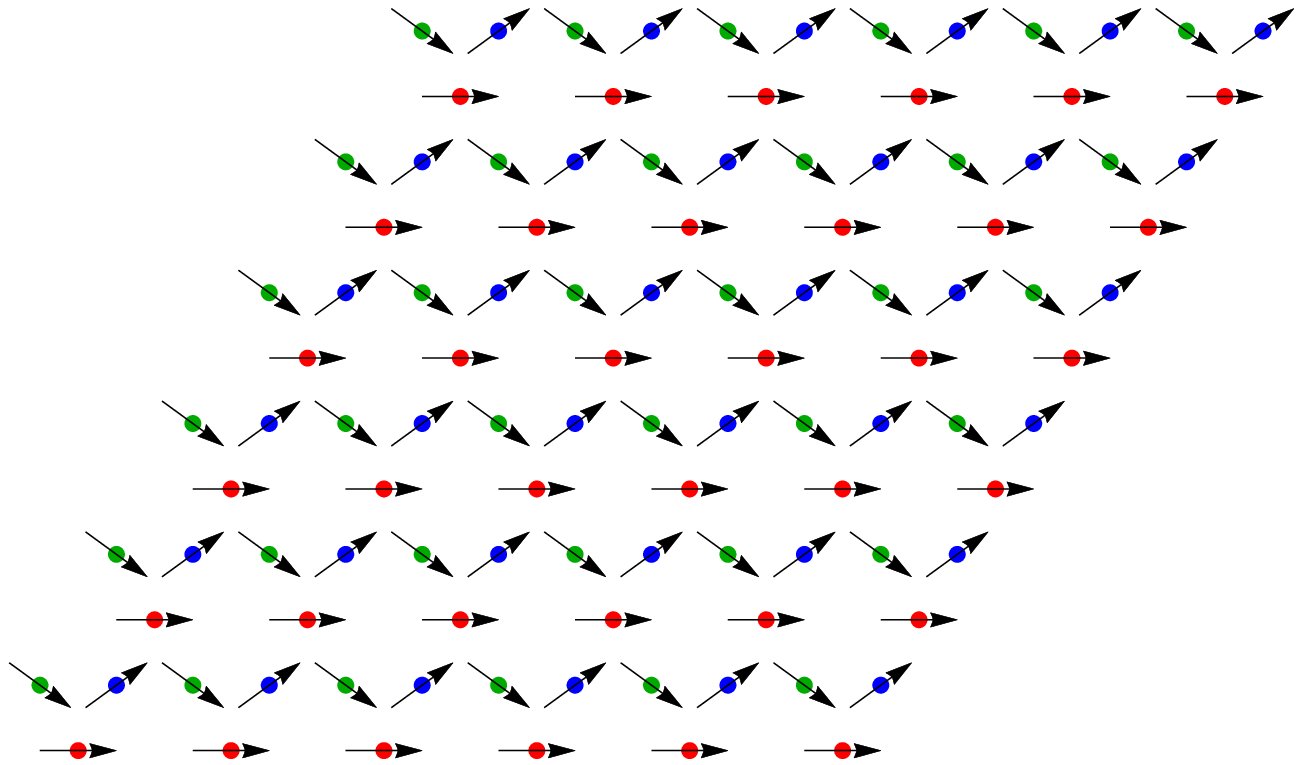}
  \caption{State 3}
 \end{subfigure}
 \begin{subfigure}{0.23\textwidth}
  \includegraphics[width=\textwidth]{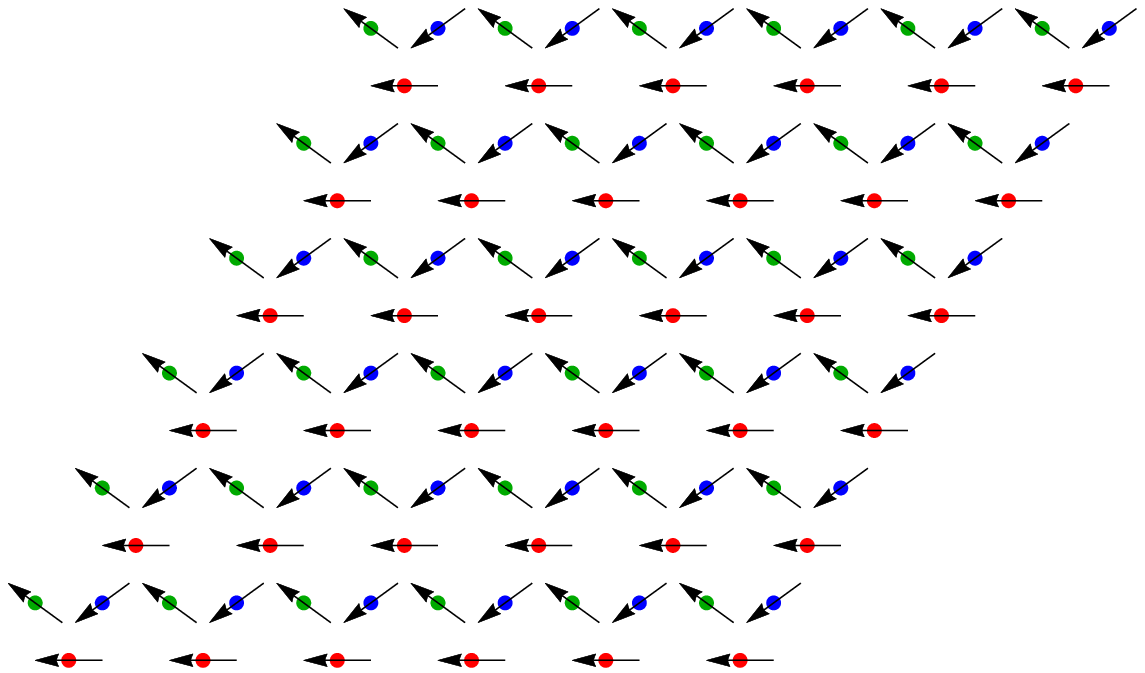}
  \caption{State 4}
 \end{subfigure}
 \begin{subfigure}{0.23\textwidth}
  \includegraphics[width=\textwidth]{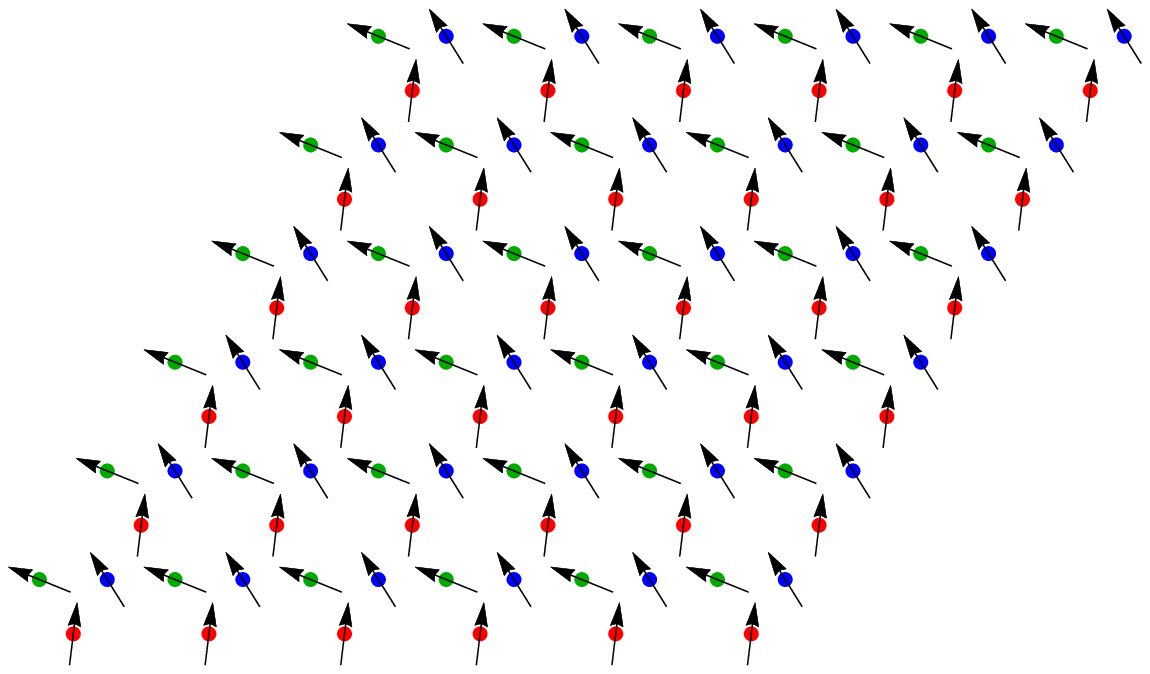}
  \caption{State 5}
 \end{subfigure}
 \begin{subfigure}{0.23\textwidth}
  \includegraphics[width=\textwidth]{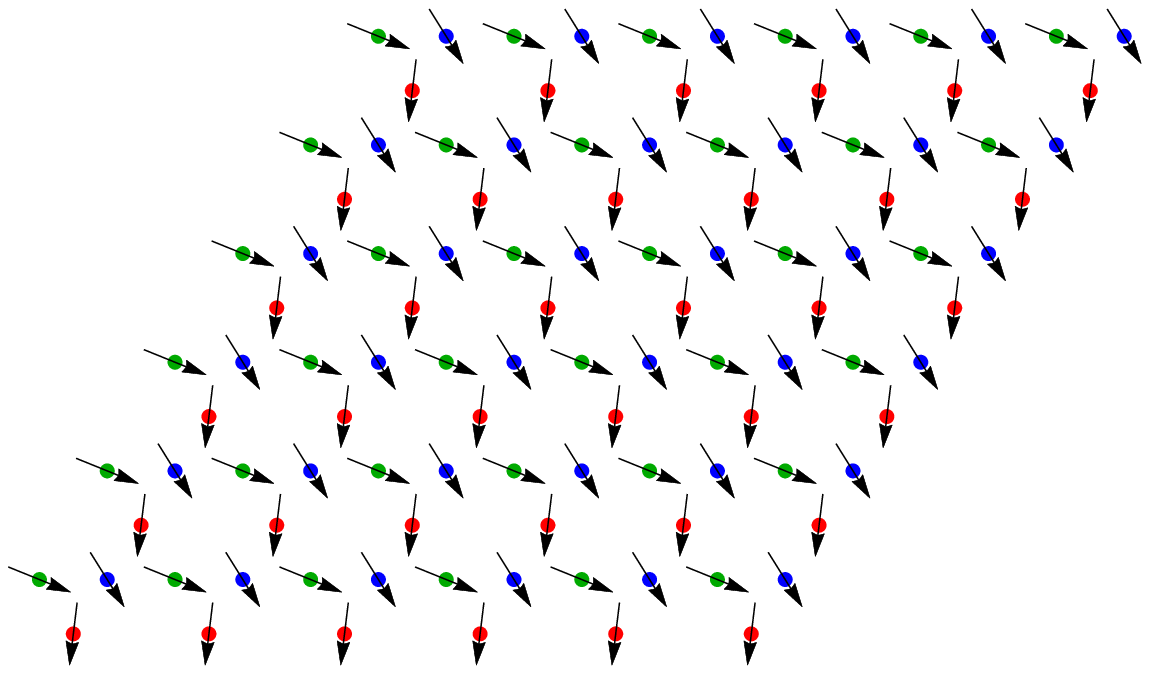}
  \caption{State 6}
 \end{subfigure}
  \caption[Six pure ground states of the 2D dipolar kagome]{Six pure ground state domains of the
2D dipolar kagome spin lattice.}
 \label{gs}
\end{figure}

This analysis reveals multi-domain spin configurations involving pairings of six types of magnetic states,
separated by domain walls as shown in Fig \ref{gssnap}. These six 
ground states, defined in Table \ref{tabgs}, all exhibit a three-sublattice structure with
fixed spin orientations for each sublattice. Each state exhibits shape anisotropy as predicted, as one sublattice in each state is aligned along the direction of $\theta = \frac{n
\pi}{3}$ relative to the horizontal axis where $n$ is an integer, $n \in \{0,1,2,3,4,5\}$. The
orientation of spins on the other two sublattices deviate from $\theta$ by an angle
$\phi=36.3887$\textdegree \hspace{1pt} such that the angles of the other two sublattices are given
by $\theta_{\pm}=\theta \pm \phi$, as seen in Table I.
 The energy of each of these six spin configurations is the same,
giving rise to a six-fold degeneracy. The six ground state spin configurations are shown in
Fig. \ref{gs}. 

\begin{table}[H]
\centering
 \begin{tabular}{|c|c|c|c|c|}
  \hline
  Domain & $\theta_A$ & $\theta_B$ & $\theta_C$ \\
  \hline
  1 & 23.6113 & 96.3887 & 60 \\
  \hline
  2 & 203.6114 & 276.3887 & 240 \\
  \hline
  3 & 36.3887 & 0 & -36.3887 \\
  \hline
  4 & 216.3887 & 180 & 143.6114 \\
  \hline
  5 & 120 & 83.6113 & 156.3887 \\
  \hline
  6 & 300 & 263.6113 & 336.3887 \\
  \hline
 \end{tabular}
 \caption{Ground states for classical dipoles on the kagome lattice. Note that the
number of the state does not correspond to the value of $n$ that defines the overall orientation
of the magnetization of that state.}
 \label{tabgs}
\end{table}

\begin{figure}[H]
\centering
 \begin{subfigure}{0.27\textwidth}
  \includegraphics[width=\textwidth]{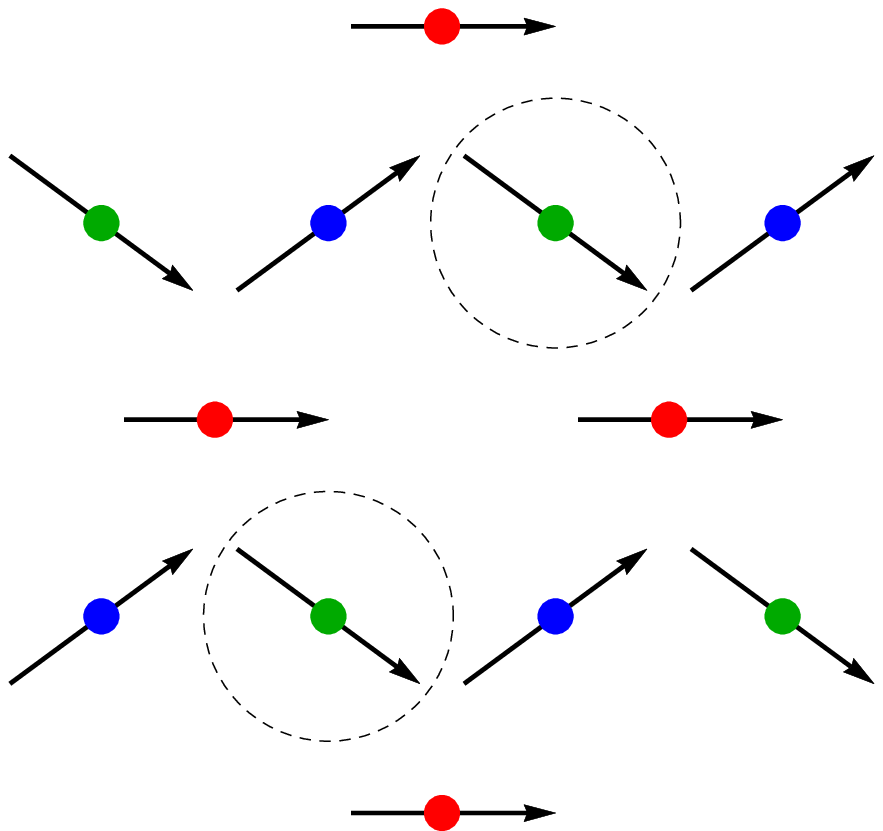}
  \caption{State 3}
 \end{subfigure}
 \begin{subfigure}{0.27\textwidth}
  \includegraphics[width=\textwidth]{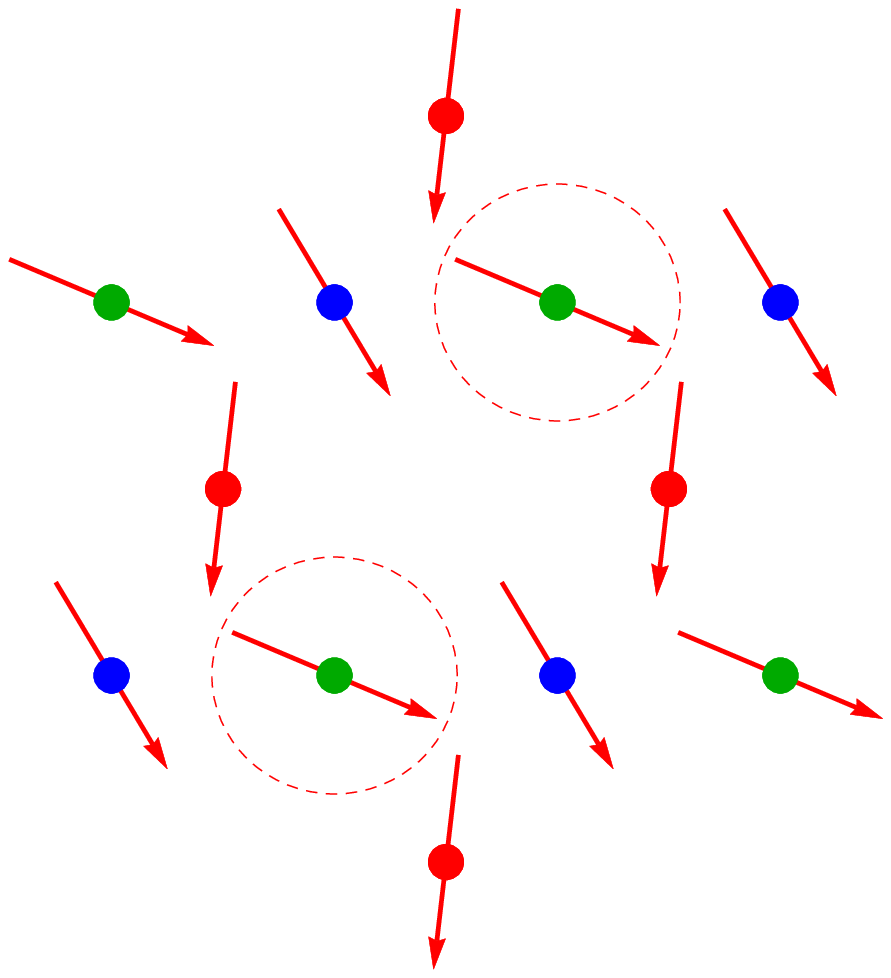}
  \caption{State 6}
 \end{subfigure}
 \begin{subfigure}{0.27\textwidth}
  \includegraphics[width=\textwidth]{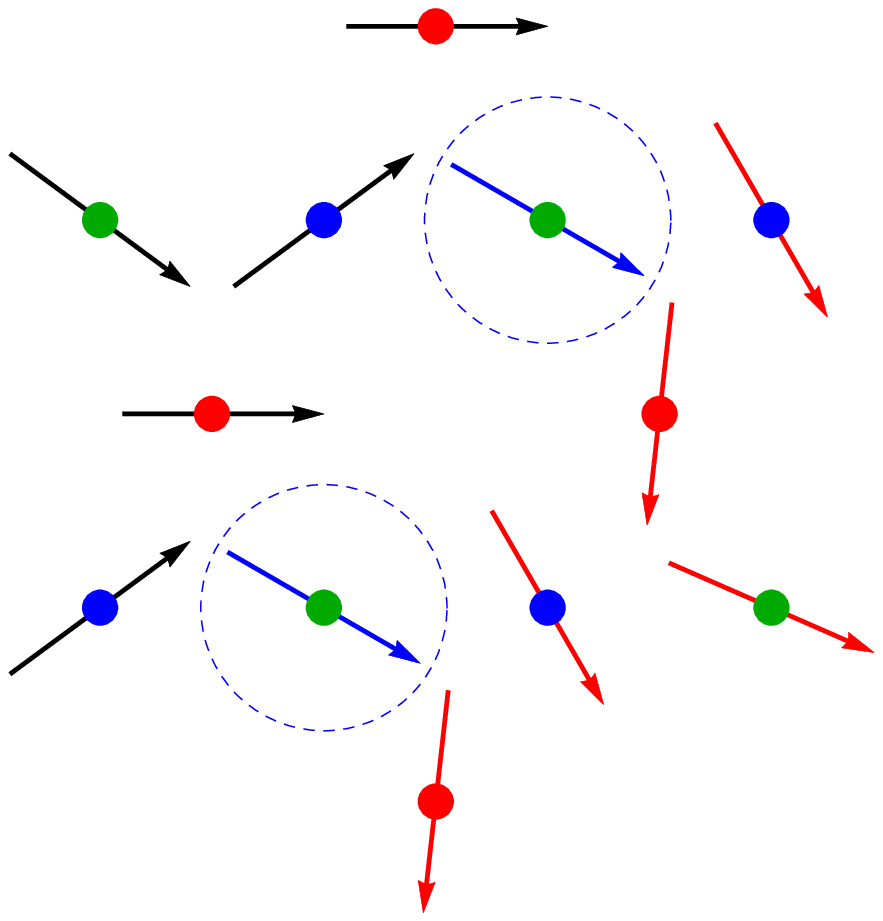}
  \caption{Mixture of States 3 and 6.}
 \end{subfigure}
 \caption{Diagram depicting domains 3 (black arrows)
and 6 (red arrows) and the domain wall (blue) formed by the mixture of these states. 
Sublattice sites are coloured blue ($A$), red ($B$) and green ($C$).
The spins
separating the two states form the domain wall with an orientation that is the average of the
orientations of the spins on sublattice $C$ (green sites) from states 3 and 6.}
 \label{sub}
\end{figure}

To further illustrate how the ground states combine to form domain walls at low
temperatures, Fig. \ref{sub} shows spin configurations 3 and 6 from Table \ref{tabgs} as pure ground states
and then shows the mixture of these two states.

The ground states have an overall orientation that can be defined by considering a
``macrospin'' formed by three spins located on a single triangle within the lattice. The overall
orientation $\theta_M$ of a domain is defined by the average orientation of the three spins that
form a macrospin, $\theta_M = (\theta + \theta_{+} +\theta_{-})/3 = \theta$. Therefore, each state
can be defined by an orientation $\theta_M = \frac{n \pi}{3}, n\in{0,1,2,3,4,5}$. Our shape
anisotropy predictions made from the magnetic dipole interaction term are confirmed further as the
spins all lie in the plane of the lattice with deviations from the plane being minimal. 

At low temperatures, the system orders itself into mixtures of these
states separated by domain walls that have a very small effect on the systems total energy. There
are six favourable pairs of states 
at low temperatures:
$(1,5)$, $(1,3)$, $(2,6)$, $(2,4)$, $(3,6)$ and $(4,5)$. Within these specific state pairings, the two participating domains have a sublattice with similar orientation, and for each of these pairs, $\theta_M$ in the two domains differs by $60^\circ$. For example, states 1 ($\theta_M = 60^\circ$) and 5 ($\theta_M = 120^\circ$)
have similar values of $\theta_B$, as seen in Table \ref{tabgs}.   The triangular lattice, which orders ferromagnetically, has less geometrical frustration than the kagome lattice, which may be responsible for the difference in ground state spin structures.

In a previous MC study on the dipolar kagome system, Tomita suggested that
the spin structure of the system at low temperatures is composed of ferromagnetic chains with spins
between the chains appearing to not be ordered\cite{tomita}. 
In fact, snapshots of the spin-structure from our simulations at temperatures studied by Tomita,
indicate distinctly different spin structures. As a result of this disagreement, we develop a
simple computational model to verify our results. By generating many configurations of the states
reported by Tomita and calculating the energy of the system, we find that the energy per spin, 
$E/{\rm spin}\approx-1.6$ to be greater than the energy per spin for our ground state,
$E/{\rm spin}=-2.38895$. 
We note that the transition temperature found by Tomita is essentially
the same as ours ($T_N \simeq 0.43$, as shown in the next section).

\begin{figure}[H]
 \centering
 \includegraphics[scale=0.55]{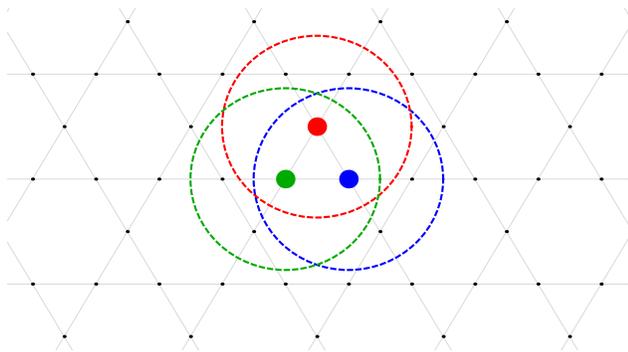}
 \caption[Visualization of finite-size model]{Visualization of the finite-size model. Spins
belonging to a central cluster (denoted by red, blue and green lattice sites) interact only with
their neighbours within the range of interaction.}
 \label{isv}
\end{figure}

Because the magnetic dipole interaction is a long-range interaction, it is of interest to study
how the ground state structure evolves as a function of the range of interaction. By considering a
finite-size model of three spins belonging to a central cluster on a kagome lattice, as in Fig.
\ref{isv}, we can study the effect of the range of the interaction on the ground state of the
system. For simplicity, we have assumed that the ground state is a three-sublattice system so that
all spins belonging to a sublattice have the same orientation. Only interactions between the central
spins and their neighbours are considered as shown in Fig. \ref{isv}. That is, the spins outside of
the central cluster do not interact with each other. The range of interaction is varied at each step
and the energy is minimized numerically in three variables, $\theta_A,\theta_B$ and $\theta_C$, the
angles of spin orientation for sublattices $A,B$ and $C$, respectively. 

\begin{figure}[H]
 \centering
 \includegraphics[scale=0.9]{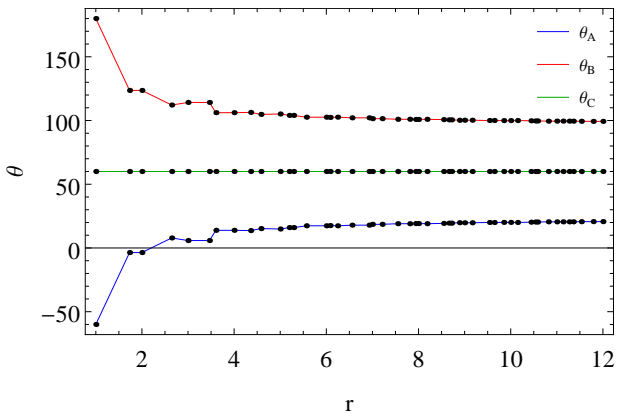}
 \caption[Sublattice angles as a function of interaction range]{Sublattice angles as a function of
the range of interaction. The angles slowly converge as the range of interaction is increased. The
ground state is nearly realized after including up to only second nearest-neighbours.}
 \label{thetatisv}
\end{figure}

The plot of the sublattice
angles as a function of the range of interaction shown in Fig. \ref{thetatisv} indicates that the underlying
physics of the magnetic dipole interaction is captured after the range includes the second
nearest-neighbours of the central spins. Interestingly, when the range only includes the
nearest-neighbours we retrieve a result that is similar to the $q=0$ ground state of the
exchange-only kagome lattice.\cite{harris} That is, the nearest-neighbour dipolar interaction gives
a  120$^\circ$ spin structure with the total magnetization on a triangle equal to
zero coupled with the shape anisotropy as all the angles point along lattice vectors. Note that in
this model, the spins in the central cluster interact with each other twice. This double counting reflects the
Ewald summations used in our simulations to calculate the correct energy per particle of the 
infinite periodic system.  

\section{SIMULATION RESULTS}
\label{sec:mc}

\begin{figure}[H]
 \centering
 \includegraphics[scale=0.92]{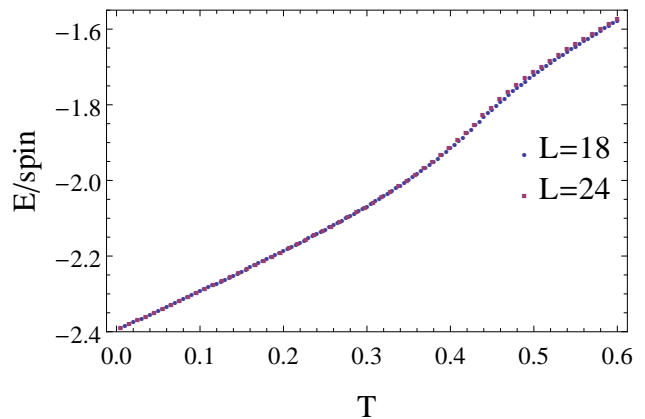}
 \caption{Energy as a function of temperature for different lattice sizes $L$ from MC cooling simulations.}
 \label{eL}
\end{figure}

Results from using Metropolis MC simulations at finite temperatures are described here.
All quantities calculated are averaged over
$10^6$ MC steps (with the initial 10\% discarded for thermalization)
unless otherwise specified and are calculated for different lattice sizes $L \times L$ with $N=3L^2/4$ spins.
Here, $L$ represents the number of points along an edge of the underlying triangular lattice, with one-quarter of the sites removed. A single MC step refers to $N$ individual Metropolis spin update attempts. 
Simulations versus temperature are done three different ways:\cite{hemmati} as cooling runs with an initial random spin configuration
and then using the final configuration of the previous temperature as the initial configuration of the
next lower value of $T$; heating runs with the initial configuration being one of the six ground states; and finally independent
temperature runs where a random initial configuration is used at each value of $T$. 
Results for the energy calculated from cooling simulations with $\Delta T \approx 0.01$ are
shown in Fig. \ref{eL} for $L$ = 18 and 24 showing little impact of system size.   There
is a small inflection in the energy at $T\approx0.4$, indicating the possibility of a phase
transition. 

\begin{figure}[H]
 \centering
 \includegraphics[scale=0.92]{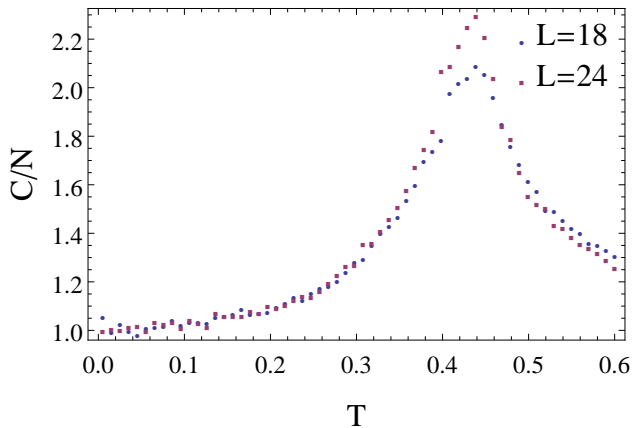}
 \caption[Specific heat vs. Temperature]{Specific heat versus temperature for different lattice
sizes $L$ from MC cooling simulations.}
 \label{cvtL}
\end{figure}

To determine if the inflection in the energy corresponds to a phase transition,
the specific heat (per spin) as a function of the temperature is calculated from cooling runs.
Fig. \ref{cvtL} shows a peak at $T\approx0.43$, consistent with Tomita\cite{tomita}, with a relatively weak dependence on system size
that may be indicative of a continuous phase transition.
  
\begin{figure}[H]
 \centering
 \includegraphics[scale=0.92]{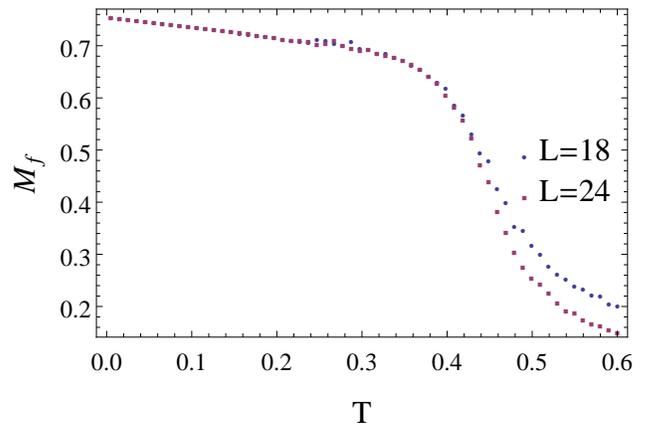}
 \caption[Total ferromagnetic magnetization versus temperature]{Total ferromagnetic magnetization
$M_f$ as a function of temperature for various lattice sizes from cooling runs.}
 \label{mfL}
\end{figure}

In classical spin systems, the total ferromagnetic magnetization,
\begin{equation} \label{mf}
 M_f=\frac{1}{N}\Bigg\langle\Bigg|{ \sum_{i} \vec{S}_i \Bigg|}\Bigg\rangle,
\end{equation}
is a useful quantity to consider as it can offer insight into the order of the system. If $M_f=0$,
the system is completely disordered with no net magnetization, while if $M_f=1$ the system is in a
ferromagnetic phase with all the spins aligned along the same orientation. Fig.
\ref{mfL} shows the total ferromagnetic magnetization from cooling runs and
indicates that the system is in a disordered phase at temperatures above  $T_N$. At low
temperatures, the results suggest that the system is in a state that exhibits order with some
deviations from fully ferromagnetic order. This is expected from the
analysis presented in Section \ref{sec:gs} as low-temperature states are composed of a mixture of
ground states  which are not totally ferromagnetic, that is $M_f \neq 1$ for a pure ground
state. In fact, for a pure ground state, $M_f=0.8700$, which is greater than the value found for an
equal mixture of two domains, for which $M_f=0.8700\, \cos{30^\circ}=0.7534$ 
(since the macrospin angles in the two domains are separated by $60^\circ$, as discussed in Section~\ref{sec:gs}).

\begin{figure}[H]
 \centering
 \includegraphics[scale=0.92]{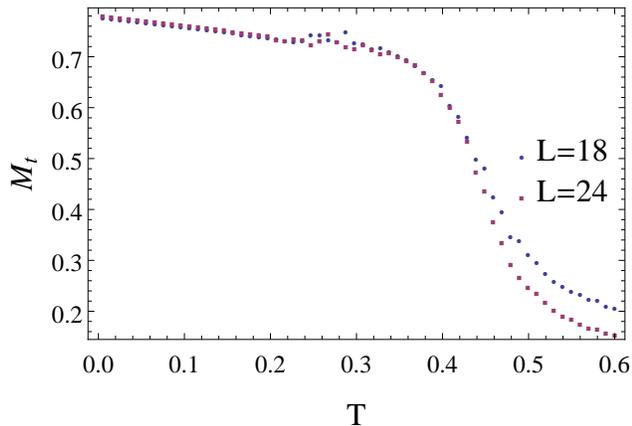}
 \caption[Total sublattice magnetization versus temperature]{Total sublattice magnetization as a
function from cooling runs. Small fluctuations at $T\approx0.3$ correspond to the system ``deciding''
which combination of domains will be present at low $T$.}
 \label{mtL}
\end{figure}

Based on the results in Section \ref{sec:gs}, we expect that each sublattice is fully ordered in the
ground state. Therefore, we calculate the sublattice magnetization order parameter,
\begin{equation} \label{mgamma}
 M_\gamma= \frac{3}{N} \Bigg\langle  \Bigg| \sum_{k \subset \gamma} \vec{S}_k \Bigg|
\Bigg\rangle,
\end{equation}
where $\gamma$ represents sublattice $A$, $B$ or $C$, and $k$ runs over all $N/3$ spins of the sublattice; as well
as the total sublattice magnetization order parameter,
\begin{equation} \label{mt}
 M_t= \frac{1}{N} \Bigg\langle \sum_{\gamma} \Bigg| \sum_{k \subset \gamma} \vec{S}_k \Bigg| \Bigg\rangle.
\end{equation}
When the value of this total sublattice magnetization is one, the system has three ferromagnetically ordered sublattices as seen
in the ground states defined in Section \ref{sec:gs}. However, at low temperatures the
presence of mixtures of ground states separated by domain walls will reduce the value of $M_t$ from
unity to a value that depends on the number of spins that belong to each domain present. Fig.
\ref{mtL} confirms this prediction as the total sublattice magnetization is not equal to unity at very
low temperatures. For a state consisting of an equal number of spins belonging to each domain,
$M_{t}=0.7757$. In Fig.~\ref{mtL}, $M_t$ tends to approximately $0.78$, which is slightly higher
than if there were and equal number of spins belonging to each domain. This is expected as the
domain with more spins will contribute more to the magnetization, bringing the value closer to
the pure ground state value of unity. Owing to the small energetic penalty of forming 
domain walls, the ground state is not realized as the appearance of a second domain will occur even at
the lowest of $T$.  The energy difference between a pure ground state of size $L=12$ and one with a single row of a complementary domain is $\Delta E = 0.38$ ($\Delta E$/spin $\approx 4\times 10^{-3}$ and
$\Delta E$ scales linearly with $L$).
This excess energy is so small that mixtures of states are inevitable in an ergodic system. 
By contrast, a single row of a non-complementary domain (e.g. state 2 with a row of state 1) yields an energetic penalty two orders of magnitude higher.

In both Fig.~\ref{mfL} and Fig.~\ref{mtL}, there exists a common feature at $T\approx 0.27$ where the
magnetization parameters appear to fluctuate. Because these fluctuations are only present for a narrow region of
temperatures, we examine the possibility that they are associated with an onset of a reduction in 
thermal-fluctation induced switching between degenerate spin states. 
We note that there is no discernable feature at this temperature region in the specific heat. 

\begin{figure}[H]
 \centering
 \includegraphics[scale=0.92]{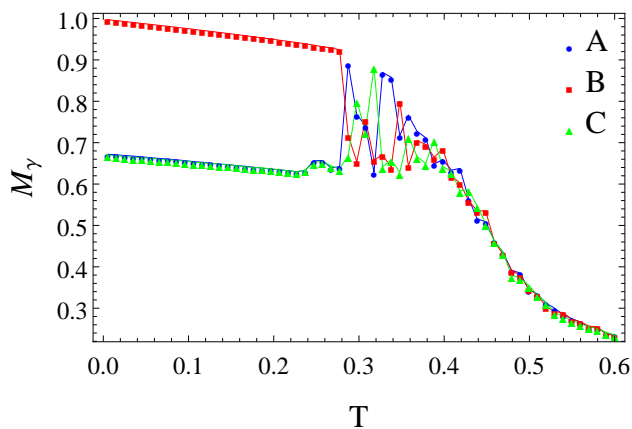}
 \caption[Total ferromagnetic magnetization for individual sublattices]{Total ferromagnetic
magnetization for individual sublattices A, B, and C from cooling runs with $L=18$.
Large fluctuations correspond to the system changing state mixtures until the system becomes
``frozen'' into a combination of states.}
 \label{subm}
\end{figure}

Fig. \ref{subm} shows that the sublattice ferromagnetic magnetizations for each sublattice of the system have large fluctuations in
the region where the features are present in Figs. \ref{mfL} and \ref{mtL}. 

\begin{figure}[H] 
 \centering
 \begin{subfigure}{0.39\textwidth}
  \includegraphics[width=\textwidth]{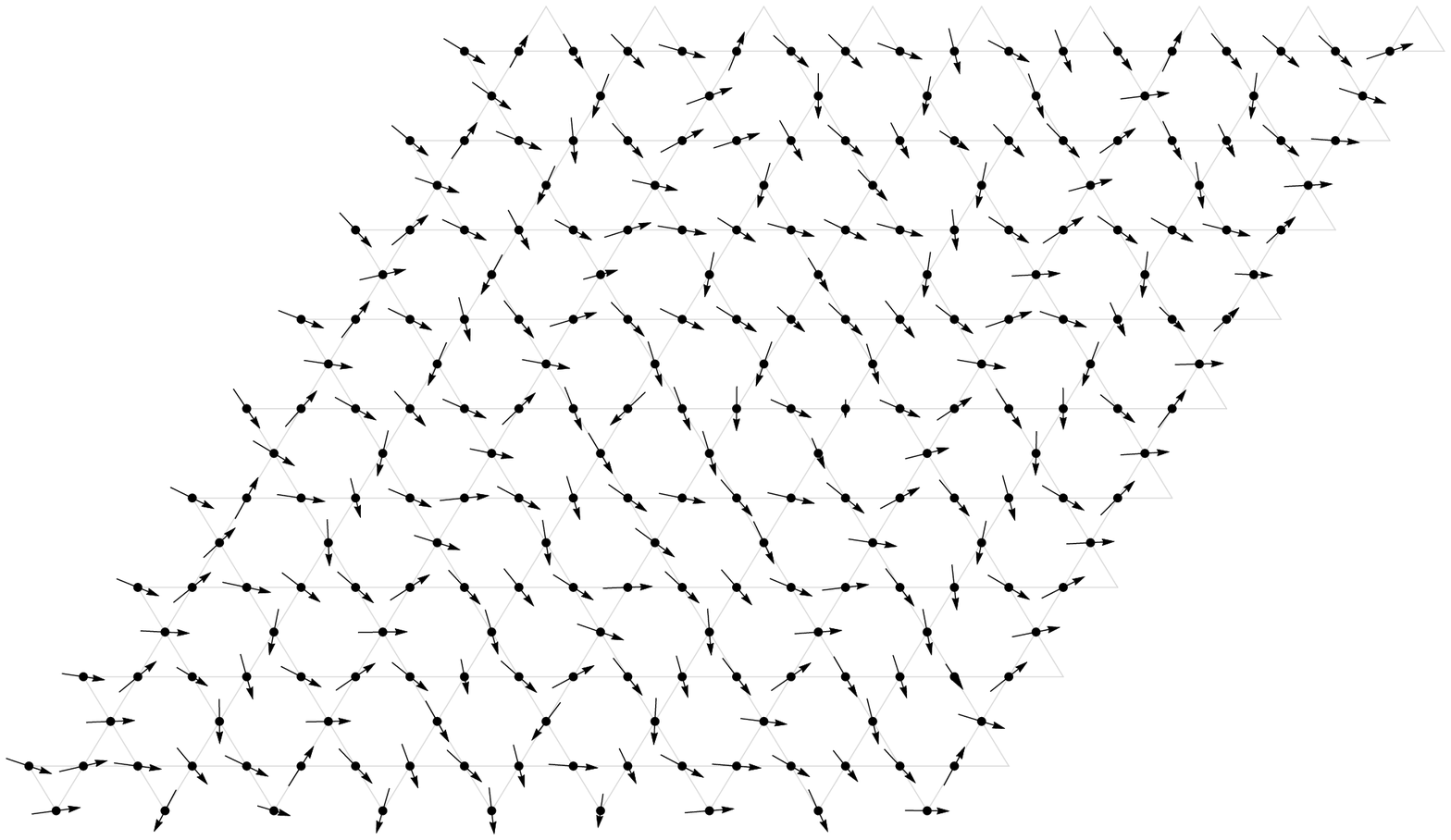}
  \includegraphics[width=\textwidth]{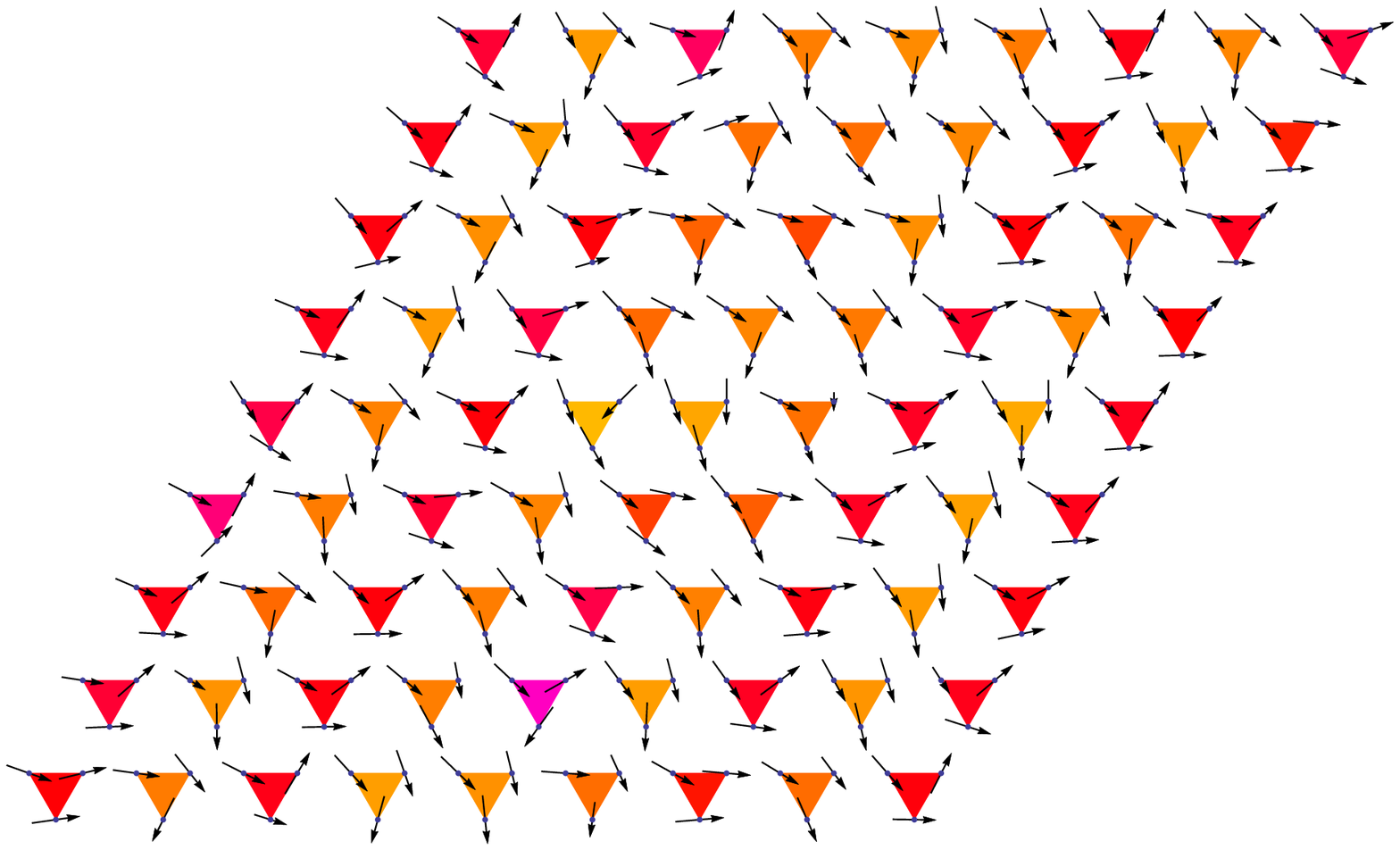}
  \caption{$T=0.3176$}
 \end{subfigure}
 \begin{subfigure}{0.39\textwidth}
  \includegraphics[width=\textwidth]{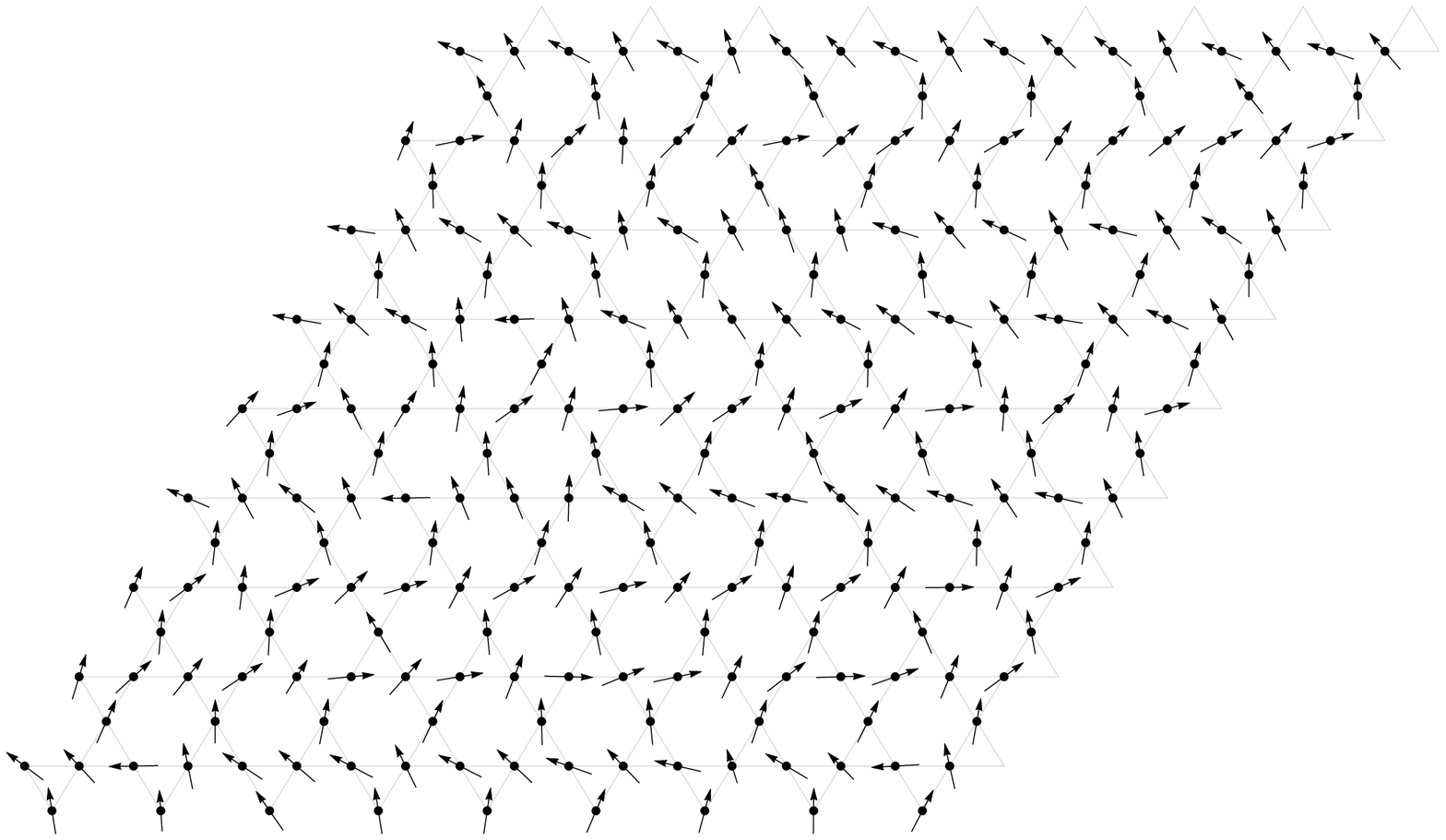}
  \includegraphics[width=\textwidth]{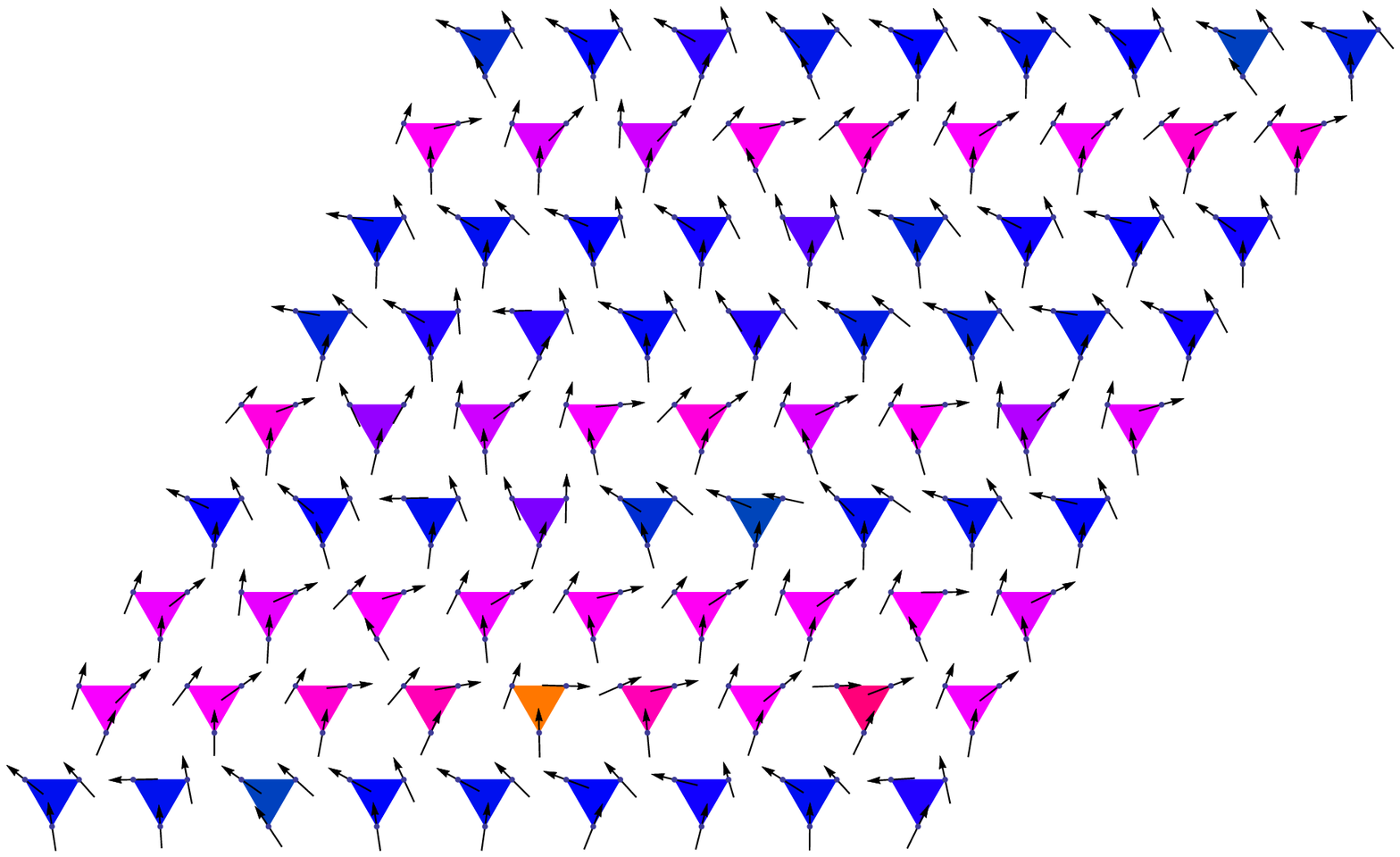}
  \caption{$T=0.2067$}
 \end{subfigure}
 \caption[Snapshots of spin structure at low temperatures]{Snapshots of the spin structure in the
region where fluctuations in the magnetization order parameters are present (See Fig. \ref{subm}).
Below the snapshots, we have included diagrams with coloured triangles used to easily identify
which states are present. Each state has an associated colour: State 1 (Magenta), State 2
(Yellow), State 3 (Red), State 4 (Green), State 5 (Blue), State 6 (Orange). The colour of each
triangle is calculated using the orientation of a macrospin $\theta_M$ as a variable in a continuous
``colour function'' that associates a Red/Green/Blue value with $\theta_M$. a) States 3 and 6 are present.
b) States 1 and 5 are present.}
 \label{snaps}
\end{figure}

By examining the spin structure in this temperature region (Fig. \ref{snaps}) the large fluctuations
can be explained by the system changing which state pairs are present. In Fig. \ref{snaps}a, a
snapshot of the MC results at a higher temperature, $T=0.3176$, shows states $3$ and $6$ are present while at
$T=0.2067$, shown in Fig \ref{snaps}b, states $1$ and $5$ are present.  For $T \gtrsim 0.3$,
the energy required to change states is low enough to allow for large fluctuations between spin configurations.
 Once the system is cooled to a sufficiently low temperature, the
energy required to change from one pair  of states to a different  pair  becomes large enough 
that the probability of the system changing
states becomes very small. At this temperature $T \lesssim 0.2$, the system becomes ``frozen'' into a
configuration that consists of a particular pair of states.

That two sublattice magnetizations are significantly less than unity and one sublattice 
magnetization is nearly at unity (as in Fig.~\ref{subm}) can be understood by 
considering Fig.~\ref{sub}, where spins on sublattice $C$ all nearly point in the same 
direction, while for sublattice $A$ (and $B$) the spins are offset by 
$96.3887^\circ$ in the two domains, giving $M_A = M_B = \cos{\left( 96.3887 / 2 \right)} = 0.6666$.

\begin{figure}[H]
 \centering
 \includegraphics[scale=0.92]{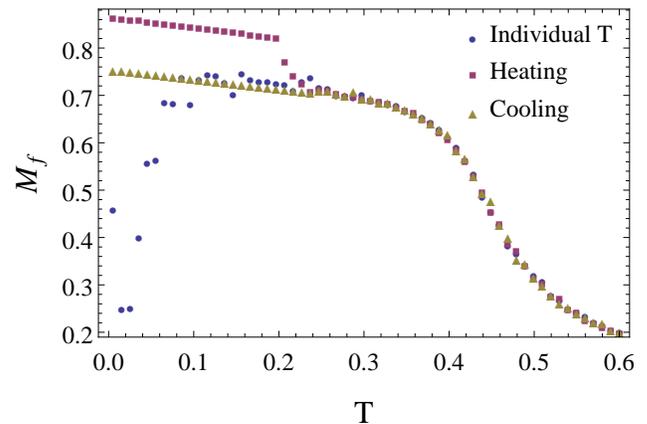}
 \caption[$M_f$ from Cooling, Heating and Individual Temperature Simulations]{Total ferromagnetic
magnetization from cooling, heating and individual 
temperature simulations on lattice of size $L=18$.}
 \label{mf_tch}
\end{figure}

In Fig. \ref{mf_tch}, results are presented for cooling, heating and individual temperature
simulations showing the total ferromagnetic magnetization as a function of temperature. Because
heating simulations start from a ground state, the ferromagnetic magnetization is higher than that
of a mixture of domains which have spins with opposing magnetizations. As the temperature is
increased, there are fluctuations introduced to the ground state spin orientations that lower the
magnetization. At $T\approx0.2$, the magnetization sharply drops to values consistent with the
cooling simulation results. Similar to the system becoming ``frozen'' into a  mixture of domains
when it is cooled to $T\approx0.2$, the system ``melts'' into a state composed of a mixture of
domains when it is heated to $T\approx0.2$. 

Also shown in Fig. \ref{mf_tch} are results from individual temperature simulations which have been
averaged over $10^7$ MC Steps. The magnetization is in agreement with both cooling and
heating simulations for temperatures $T>0.2$ but has a large amount of noise for temperatures below
$T\approx0.2$. This noise is resultant from the fact that at each temperature studied the system
has to be equilibrated from a randomly generated initial state. Since the acceptance probability
is proportional to the Boltzmann factor,  as the temperature is decreased the
acceptance probability will decrease exponentially. Therefore, at low temperatures
the number of Monte Carlo Steps required to equilibrate the system increases greatly which
increases the computational time required to perform a simulation that yields good statistics.

\begin{figure}[H]
 \centering
 \includegraphics[scale=0.92]{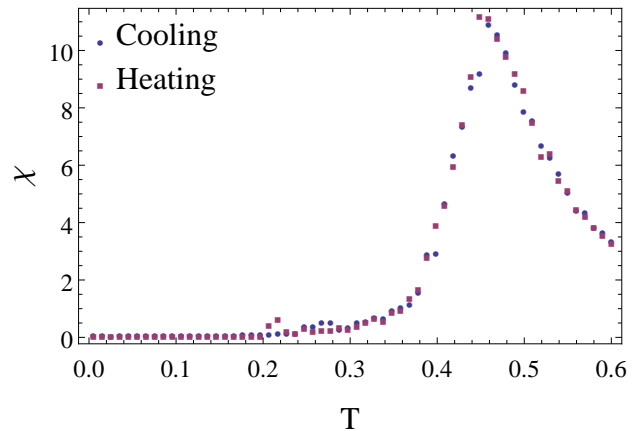}
 \caption[Magnetic susceptibility from cooling and heating runs simulations]{Magnetic Susceptibility
as a function of temperature for cooling and 
heating simulations on a lattice of size $L=18$.}
 \label{chi_ch}
\end{figure}

Finally, Fig. \ref{chi_ch} shows the magnetic susceptibility\cite{leblanc1} as a function of 
temperature for cooling and heating simulations. Both simulations yield a peak at $T\approx0.43$ as 
expected from previous results that identified a phase transition at this temperature. The results 
also identify features corresponding to how the system changes as it is heated or cooled. From the 
heating results, we see a small peak in the susceptibility at $T\approx0.2$ where the system begins 
to form domain walls. Cooling simulation results show that the susceptibility experiences 
fluctuations at a temperature range close to $T\approx0.3$ which results from the system reordering 
into mixtures of domains
during this range.

\section{SUMMARY AND CONCLUSIONS}
\label{sec:conclusions}
As a prelude to studies of magnetic thin-film geometries which mimic spin-valve exchange bias structures, and as a compliment to studies
of magnetic nano-structures,
classical Heisenberg spins with only dipole-dipole interactions on the 2D kagome lattice have been studied to identify the
ground state spin structures, degeneracies and thermal behaviour of this geometrically frustrated lattice.
In contrast with the simple triangular lattice with edge-sharing triangles which exhibits a ferromagnetic ground state,
our EFM and low-T MC simulations reveal that the corner-sharing kagome structure leads to a more complicated six-fold degenerate 
spin states.  These have been characterized (see Table I) in terms of three
ferromagnetically ordered sublattices and multi-state domain structures that result from the simulations have been identified.
Similarities and differences from the near-neighbor antiferromagnetic exchange only 120$^\circ$ spin structures of the triangular and kagome lattices are described and result from beyond-third-neighbor dipolar terms where shape anisotropy effects become more important.

The thermal behaviour of the system was studied through cooling, heating and individual
temperature MC simulations. Results suggest that the system undergoes a phase transition
at $T\approx0.43$ in agreement with previous MC simulations\cite{tomita} but the nature of the ordered state differs.
Ferromagnetic and sublattice magnetization order parameters calculated
during cooling simulations suggest that below the critical temperature, the system changes between
mixtures of ground state domains with fluctuations present until a threshold temperature
$T\approx0.2$ is reached where the system becomes ``frozen'' into a specific mixture of ground state
domains. Heating simulations yield corroborating results in that the system has only one domain present
at low temperatures until it reaches $T\approx0.2$ where the system readily changes into a mixture
of domains as reflected in Fig. \ref{mf_tch}. Our results suggest that 
the phase trasnition at $T_N$ is continuous and its criticality is likely impacted by domain fluctuations
and the long-range nature of the dipole interaction.\cite{frey}

Also of interest is to extend this analysis to study dipole interactions on the 3D fcc ABC-stacked kagome systems.\cite{hemmati} 
Preliminary results\cite{holden} indicate a phase transition to long range order at $T_N \sim 0.4$ to a state with a different
spin structure from the 2D kagome system and also different from the ferromagnetic order in the standard fcc lattice.\cite{bouchard}

Finally we note that during the preparation of this manuscript we became aware of work that shows similar six-fold degenerate ground states as described here.\cite{maksy}

This work was supported by the Natural Sciences and Engineering Research Council (NSERC) of Canada,
and the Compute Canada facilities of the Atlantic Computational Excellence network (ACEnet) and the Western Canada Research Grid (WestGrid).

\end{document}